\documentclass[preprint,amsmath,amssymb,aps,showkeys,showpacs]{revtex4}
\usepackage[english]{babel}
\usepackage{graphicx}
\usepackage{graphics}
\usepackage{amsmath}
\usepackage{dcolumn}
\usepackage{amssymb}
\usepackage{bm}
\usepackage[latin1]{inputenc}

\begin{document}
\title{Quantum Holonomies based on the Lorentz-violating tensor background}
\author{K. Bakke}
\email{kbakke@fisica.ufpb.br}
\affiliation{Departamento de F\'isica, Universidade Federal da Para\'iba, Caixa Postal 5008, 58051-970, Jo\~ao Pessoa, PB, Brazil.}

\author{H. Belich} 
\affiliation{Departamento de F\'isica e Qu\'imica, Universidade Federal do Esp\'irito Santo, Av. Fernando Ferrari, 514, Goiabeiras, 29060-900, Vit\'oria, ES, Brazil.}

\begin{abstract}
We study geometric quantum phases corresponding to analogues of the Anandan quantum phase [J. Anandan, Phys. Lett. A {\bf138}, 347 (1989)] based on a possible scenario of the Lorentz symmetry violation background in a tensor background. We also show that quantum holonomies associated with the analogue of the Anandan quantum phase can be determined, and discuss a way of performing one-qubit quantum gates by analogy with the holonomic quantum computation [P. Zanardi and M. Rasetti, Phys.  Lett. A {\bf264}, 94 (1999)].

\end{abstract}

\maketitle

\section{Introduction}

Symmetry breaking is well-known in nonrelativistic quantum systems involving phase transitions such as ferromagnetic systems, where the rotation symmetry is broken when the system is under the influence of a magnetic field. Similarly, in superconductors the spontaneous violation of gauge symmetry shields the electromagnetic interaction, but in type II superconductors the magnetic field penetrates as determined by Abrikosov \cite{ANO}, forming a 2D vortice lattice.

For relativistic systems, the study of symmetry breaking can be extended by considering a background given by spacetime indices of tensor with rank $n\geq 1$. The background field, in this situation, breaks the symmetry $SO\left(1,3\right)$ instead of the symmetry $SO\left(3\right)$. This line of research is known in the literature as the spontaneous violation of the Lorentz symmetry \cite{extra3,extra1,extra2}. This new possibility of spontaneous violation was first suggested in 1989 in a work of Kostelecky and Samuel \cite{extra3} indicating that, in the string field theory, the spontaneous violation of symmetry by a scalar field could be extended. This extension has as immediate consequence: a spontaneous breaking of the Lorentz symmetry. In the electroweak theory, a scalar field acquires a nonzero vacuum expectation value which yields mass to gauge bosons (Higgs Mechanism). Similarly, in the string field theory, this scalar field can be extended to a tensor field. Nowadays, these theories are encompassed in the framework of the Standard Model Extension (SME) \cite{col} as a possible extension of the minimal Standard Model of the fundamental interactions. For instance, the violation of the Lorentz symmetry is implemented in the fermion section of the Standard Model Extension by two CPT-odd terms: $a_{\mu}\overline{\psi}\gamma^{\mu}\psi$ and $b_{\mu}\overline{\psi}\gamma_{5}\gamma^{\mu}\psi$, where $a_{\mu}$ and $b_{\mu}$ correspond to the Lorentz-violating vector backgrounds. From these fixed vector field backgrounds, Lorentz symmetry breaking effects have been investigated in quantum Hall effect \cite{lin2}, bound states solutions \cite{bel,belich2,bbs,bb,bb2} and geometric quantum phases \cite{belich,belich1,belich3,bbs2,bbs3}.

The CPT-even gauge sector of SME is obtained by including the gauge sector in the Lagrangian term, $-\frac{1}{4}\left(k_{F}\right)_{\mu\nu\kappa\lambda}\,F^{\mu\nu}\left(x\right)F^{\kappa\lambda}\left(x\right)$, with $\left(k_{F}\right)_{\mu\nu\kappa\lambda}$ being a Lorentz-violating tensor. This tensor is composed of 19 coefficients, where nine of these coefficients are nonbirefringent and ten are birefringent, being all of them endowed with the symmetries of the Riemann tensor and a double null trace ($\left(k_{F}\right)^{\mu\nu}_{\,\,\,\,\,\,\,\mu\nu}=0$). The effects of this CPT-even electrodynamics on the fermion-fermion interaction was considered in Refs. \cite{18}.

Our interest in this work is to study the appearance of geometric quantum phases that stem from a Lorentz-violating tensor background. The arising of geometric quantum phases in interferometry experiments stems from the presence of a potential vector along the path of a charged particle even though there exists no interaction with a magnetic field \cite{ab} or electric field \cite{dab,dab2}. Geometric phases was introduced by Berry \cite{berry} in 1984 to describe the phase shift acquired by the wave function of a quantum particle in an adiabatic cyclic evolution. At present days, it is well-known that geometric quantum phases can be measured in any cyclic evolution \cite{ahan,anan4,anan5}. The best famous quantum effect related to the appearance of geometric phases is the Aharonov-Bohm effect (AB) \cite{ab}. It worth mentioning other quantum effects related to geometric phases that are termed the dual effect of the Aharonov-Bohm effect \cite{dab,dab2} and the scalar Aharonov-Bohm \cite{pesk}. 

In recent years, the study of geometric quantum phases has been extended to neutral particles with permanent magnetic dipole moment \cite{ac} and permanent electric dipole moment \cite{hmw,hmw2}. However, the quantum effects associated with geometric phases for neutral particles stem from the interaction between the magnetic (electric) dipole moment of the neutral particle with electric (magnetic) field and are considered an AB-type effect in the sense that this interaction is a force-free interaction \cite{disp3}. Well-known quantum effects associated with geometric phases for neutral particle are the Aharonov-Casher effect \cite{ac}, the He-McKellar-Wilkens effect \cite{hmw,hmw2}, the scalar AB effect for neutral particles \cite{zei,anan,anan2}. Recently, analogues effects for neutral particle have been studied, such as, an analogue of the He-McKellar-Wilkens effect \cite{furt,bbs3}, analogues of the Aharonov-Casher effect \cite{furt2,bf12,bbs2} and analogues of the scalar AB effect for neutral particles \cite{bf15,bf30,bbs3,b25}.

In this paper, we show that analogues of the Anandan quantum phase \cite{anan,anan2} can be obtained by defining possible scenarios of the Lorentz symmetry breaking induced by tensor background. Moreover, we show that quantum holonomies associated with the analogue of the Anandan quantum phase can be determined and, by analogy with the holonomic quantum computation \cite{zr1,vedral,vedral2}, we discuss a way of performing one-qubit quantum gates for a Dirac neutral particle.

This paper is organized as follows: in section II, we introduce the Lorentz symmetry violation background and discuss the nonrelativistic limit of the Dirac equation for a neutral particle; in section III, we obtain an analogue of the Anandan quantum phase \cite{anan,anan2} and discuss the analogy with the holonomic quantum computation \cite{zr1,vedral,vedral2}; in section IV, we present our conclusions.

\section{Tensor Background of the Lorentz symmetry violation}

In this section, we introduce the Lorentz symmetry violation background and discuss the nonrelativistic limit of the Dirac equation for a neutral particle. The scenario of the Lorentz symmetry breaking is based on a fixed tensor field by modifying a fermionic coupling studied in Ref. \cite{belich}. This nonminimal coupling corresponds to corrections of a dynamical field which goes on to present a new behaviour when the physical system begins to access a new energy scale. Such nonminimal coupling with the background can present relevant information about the low energy regime of a new theory. Hence, if we have a fundamental theory with vector or tensor fields violating the Lorentz symmetry, by assuming non-trivial expected values in vacuum, we propose that the fermion sector should feel this background by a nonminimal coupling with this background. Specifically, we present it as a coupling $\frac{\lambda _{1}}{2}\,H_{\mu\nu}\,\,\Sigma^{\mu\nu}$ and $\lambda _{2}\,H_{\mu\alpha}\,F_{\,\,\,\nu}^{\alpha}\left(x\right)\,\Sigma ^{\mu\nu}$. We deal with an effective theory in a background that violates the
Lorentz symmetry describing the behaviour of a neutral Fermion. Therefore, we consider a natural particle which describes this behaviour is a neutron moving in this background. A practical experimental configuration for testing the geometric phases studied in this work is a neutron beam interferometer. Thereby, we write the Dirac equation in the following form:
\begin{eqnarray}
m\psi=i\gamma^{\mu}\partial_{\mu}\psi+\frac{\lambda_{1}}{2}\,H_{\mu\nu}\,\Sigma^{\mu\nu}\psi+\lambda_{2}\,H_{\mu\alpha}\,F^{\alpha}_{\,\,\,\nu}\left(x\right)\,\Sigma^{\mu\nu}\psi,
\label{2.1}
\end{eqnarray}
where the element that gives rise to the Lorentz symmetry violation is a skew-symmetric tensor $H_{\mu\nu}$. From the properties of a skew-symmetric tensor, we define two vectors $\vec{T}$ and $\vec{S}$ from the tensor $H_{\mu\nu}$ in the following form:
\begin{eqnarray}
H_{0i}=T_{i};\,\,\,\,\,H_{ij}=\epsilon_{ijk}\,S^{k}.
\label{2.2}
\end{eqnarray}
Furthermore, the tensor $F_{\mu\nu}\left(x\right)$ in (\ref{2.1}) corresponds to the usual electromagnetic tensor ($F_{0i}=-F_{i0}=E_{i}$, and $F_{ij}=-F_{ji}=\epsilon_{ijk}B^{k}$), $\Sigma^{ab}=\frac{i}{2}\left[\gamma^{a},\gamma^{b}\right]$ and the $\gamma^{\mu}$ matrices are defined in the Minkowski spacetime in the form \cite{greiner}:
\begin{eqnarray}
\gamma^{0}=\hat{\beta}=\left(
\begin{array}{cc}
1 & 0 \\
0 & -1 \\
\end{array}\right);\,\,\,\,\,\,
\gamma^{i}=\hat{\beta}\,\hat{\alpha}^{i}=\left(
\begin{array}{cc}
 0 & \sigma^{i} \\
-\sigma^{i} & 0 \\
\end{array}\right);\,\,\,\,\,\,\Sigma^{i}=\left(
\begin{array}{cc}
\sigma^{i} & 0 \\
0 & \sigma^{i} \\	
\end{array}\right),
\label{2.3}
\end{eqnarray}
with $\vec{\Sigma}$ being the spin vector. The matrices $\sigma^{i}$ correspond to the standard Pauli matrices that satisfy the relation $\left(\sigma^{i}\,\sigma^{j}+\sigma^{j}\,\sigma^{i}\right)=2\eta^{ij}$, where $\eta_{\mu\nu}=\mathrm{diag}\left(-\,+\,+\,+\right)$ is the Minkowski tensor. 

In this work, we deal with a general coordinate system, that is, we intend to work with curvilinear coordinates, then, we apply a coordinate transformation $\frac{\partial}{\partial x^{\mu}}=\frac{\partial \bar{x}^{\nu}}{\partial x^{\mu}}\,\frac{\partial}{\partial\bar{x}^{\nu}}$, and a unitary transformation on the wave function $\psi\left(x\right)=U\,\psi'\left(\bar{x}\right)$ \cite{schu,bbs3,bb2,bbs}. Thereby, the Dirac equation can be written in any orthogonal system in the presence of Lorentz symmetry breaking effects described in (\ref{2.1}) as \cite{bbs,bbs3,bb2} 
\begin{eqnarray}
i\,\gamma^{\mu}\,D_{\mu}\,\psi+\frac{i}{2}\,\sum_{k=1}^{3}\,\gamma^{k}\,\left[D_{k}\,\ln\left(\frac{h_{1}\,h_{2}\,h_{3}}{h_{k}}\right)\right]\psi+\frac{\lambda_{1}}{2}\,H_{\mu\nu}\,\Sigma^{\mu\nu}\psi+\lambda_{2}\,H_{\mu\alpha}\,F^{\alpha}_{\,\,\,\nu}\left(x\right)\,\Sigma^{\mu\nu}\psi=m\psi,
\label{2.4}
\end{eqnarray}
where $D_{\mu}=\frac{1}{h_{\mu}}\,\partial_{\mu}$ is the derivative of the corresponding coordinate system, and the parameter $h_{k}$ correspond to the scale factors of this coordinate system \cite{schu}. For instance, the line element of the Minkowski spacetime is writing in cylindrical coordinates in the form: $ds^{2}=-dt^{2}+d\rho^{2}+\rho^{2}d\varphi^{2}+dz^{2}$; then, the corresponding scale factors are $h_{0}=1$, $h_{1}=1$, $h_{2}=\rho$ and $h_{3}=1$. Moreover, the second term in (\ref{2.4}) gives rise to a term called the spinorial connection $\Gamma_{\mu}\left(x\right)$ \cite{schu,bbs,bbs2,bbs3,weinberg}. In this way, the Dirac equation (for instance, in cylindrical coordinates) in the Lorentz symmetry violation background described by the nonminimal coupling (\ref{2.1}) is given by \cite{bbs3,bbs}
\begin{eqnarray}
m\psi=i\gamma^{\mu}\partial_{\mu}\psi+i\gamma^{\mu}\,\Gamma_{\mu}\left(x\right)\,\psi+i\lambda_{1}\vec{\alpha}\cdot\vec{T}\psi+i\lambda_{2}\,\vec{\alpha}\cdot\vec{\mathbb{A}}\psi+\lambda_{1}\vec{\Sigma}\cdot\vec{S}\psi+\lambda_{2}\vec{\Sigma}\cdot\vec{\mathbb{G}}\,\psi,
\label{2.5}
\end{eqnarray}
where we have written $i\gamma^{\mu}\,\Gamma_{\mu}\left(x\right)=\frac{i}{2}\,\sum_{k=1}^{3}\,\gamma^{k}\,\left[D_{k}\,\ln\left(\frac{h_{1}\,h_{2}\,h_{3}}{h_{k}}\right)\right]$, and defined the effective fields $\vec{\mathbb{A}}$ and $\vec{\mathbb{G}}$ in the Dirac equation (\ref{2.5}) as
\begin{eqnarray}
\vec{\mathbb{A}}&=&-\left(\vec{T}\times\vec{B}\right)+\left(\vec{E}\times\vec{S}\right)\nonumber\\
[-2mm]\label{2.6}\\[-2mm]
\vec{\mathbb{G}}&=&\left(\vec{T}\times\vec{E}\right)+\left(\vec{B}\times\vec{S}\right).\nonumber
\end{eqnarray}

In this work, we are interested in discussing the nonrelativistic behavior of a Dirac neutral particle in a background of the Lorentz symmetry violation described by (\ref{2.1}). In this study, we apply the Foldy-Wouthuyssen approximation \cite{fw,greiner} up to the terms of order $m^{-1}$. In this approach, we need first to write the Dirac equation in the form:
\begin{eqnarray}
i\frac{\partial\psi}{\partial t}=\hat{H}\psi,
\label{4.1}
\end{eqnarray}
where the Hamiltonian of the system must be write as a linear combination of even terms $\hat{\epsilon}$ and odd terms $\hat{O}$ as $\hat{H}=\hat{\beta}\,m+\hat{O}+\hat{\epsilon}$, where the operators $\hat{\mathcal{O}}$ and $\hat{\mathcal{E}}$ satisfy the relations:
\begin{eqnarray}
\hat{\mathcal{O}}\hat{\beta}+\hat{\beta}\hat{\mathcal{O}}=0;\,\,\,\,\,\,\,\hat{\mathcal{E}}\hat{\beta}-\hat{\beta}\hat{\mathcal{E}}=0.
\label{2.9}
\end{eqnarray}
After few steps, by considering just terms up to the order of $m^{-1}$, the nonrelativistic limit of the Dirac equation is given by
\begin{eqnarray}
i\frac{\partial\psi}{\partial t}=m\hat{\beta}\psi+\hat{\mathcal{E}}\psi+\frac{\hat{\beta}}{2m}\,\hat{\mathcal{O}}^{2}\psi.
\label{2.8}
\end{eqnarray}

Returning to the Dirac equation (\ref{2.5}), in order to apply the Foldy-Wouthuyssen approach \cite{fw,greiner}, we rewrite (\ref{2.5}) in the form:
\begin{eqnarray}
i\frac{\partial\psi}{\partial t}=m\hat{\beta}\psi+\vec{\alpha}\cdot\vec{\pi}\,\psi-i\lambda_{1}\,\hat{\beta}\,\vec{\alpha}\cdot\vec{T}\psi-i\,\lambda_{2}\,\hat{\beta}\,\vec{\alpha}\cdot\vec{\mathbb{A}}\,\psi-\lambda_{1}\,\hat{\beta}\,\vec{\Sigma}\cdot\vec{S}\,\psi-\lambda_{2}\,\hat{\beta}\,\vec{\Sigma}\cdot\vec{\mathbb{G}}\,\psi,
\label{2.7}
\end{eqnarray}
where we have written $\vec{\pi}=\vec{p}-i\vec{\xi}$ and $-i\xi_{k}=-\frac{1}{2\rho}\,\sigma^{3}\,\delta_{2k}$ \cite{bbs3,bb2,bb}. From Eq. (\ref{2.7}), the operators $\hat{\mathcal{O}}$ and $\hat{\mathcal{E}}$ which satisfy (\ref{2.9}) are
\begin{eqnarray}
\hat{\mathcal{O}}&=&\vec{\alpha}\cdot\vec{\pi}-i\lambda_{1}\hat{\beta}\vec{\alpha}\cdot\vec{T}-i\lambda_{2}\hat{\beta}\,\vec{\alpha}\cdot\vec{\mathbb{A}}
;\nonumber\\
[-2mm]\label{2.10}\\[-2mm]
\hat{\mathcal{E}}&=&-\lambda_{1}\hat{\beta}\,\vec{\Sigma}\cdot\vec{S}-\lambda_{2}\hat{\beta}\,\vec{\Sigma}\cdot\vec{\mathbb{G}}.\nonumber
\end{eqnarray}

Substituting (\ref{2.10}) in (\ref{2.8}), we can write the nonrelativistic limit of the Dirac equation (\ref{2.7}) in the form (for two-spinors):
\begin{eqnarray}
i\frac{\partial\psi}{\partial t}&=&m\psi+\frac{1}{2m}\left[\vec{p}-i\vec{\xi}+\lambda_{1}\left(\vec{\sigma}\times\vec{T}\right)+\lambda_{2}\,\left(\vec{\sigma}\times\vec{\mathbb{A}}\right)\right]^{2}\psi+\frac{\lambda_{1}}{2m}\left(\vec{\nabla}\cdot\vec{T}\right)\,\psi\nonumber\\
[-2mm]\label{2.11}\\[-2mm]
&+&\frac{\lambda_{2}}{2m}\left(\vec{\nabla}\cdot\vec{\mathbb{A}}\right)\,\psi-\frac{\lambda_{2}^{2}}{2m}\,\mathbb{A}^{2}\,\psi-\frac{\lambda_{1}^{2}}{2m}\,T^{2}\,\psi-\lambda_{1}\,\vec{\sigma}\cdot\vec{S}\psi-\lambda_{2}\,\vec{\sigma}\cdot\vec{\mathbb{G}}\,\psi,\nonumber
\end{eqnarray}  

Note that the first term of the right-hand side of (\ref{2.11}) corresponds to the rest energy of the nonrelativistic Dirac neutral particle \cite{greiner}. The remaining terms of the right-hand side of (\ref{2.11}) correspond to the Schr\"odinger-Pauli equation based on a Lorentz symmetry breaking scenario defined by a tensor background $H_{\mu\nu}$.  Recently, an Abelian geometric phase corresponding to an analogue of the Aharonov-Casher effect \cite{ac} has been obtained in \cite{belich}. In this work, we show that an analogue of the Aharonov-Casher effect \cite{ac} cannot be achieved, but an analogue of the Anandan quantum phase \cite{anan,anan2} can be obtained.

It worth mentioning that a similar method is presented in Ref. \cite{math} for deriving the nonrelativistic quantum Hamiltonian operator of a free massive fermion from a relativistic Hamiltonian operator with Lorentz-violating terms. The authors focus on terms from the Standard Model Extension. In our case, we are interested in obtaining a phase shift that stems from the interaction between the spin of a neutral particle and electromagnetic fields, that is, we are interested in configurations of different couplings with the Lorentz symmetry violation background.

\section{Quantum holonomies induced by the Lorentz-violating tensor background}

We start this section by discussing the appearance of geometric quantum phases in the wave function of a nonrelativistic Dirac neutral particle induced by a Lorentz symmetry breaking scenario defined by the Lorentz-violating tensor background $H_{\mu\nu}$. We show that this possible scenario of the Lorentz symmetry breaking induces an analogue of the Anandan quantum phase \cite{anan,anan2}. In Refs. \cite{anan,anan2}, Anandan showed that a geometric quantum phase arises from the interaction between the permanent magnetic dipole moment $\vec{\mu}=\mu\,\vec{\sigma}$ of a Dirac neutral particle and the electromagnetic field. This geometric phase is yielded by the presence of an effective potential vector $b_{\mu}=\left(-\vec{\sigma}\cdot\vec{B},\,\vec{\sigma}\times\vec{E}\right)$, were $\vec{\sigma}$ corresponds to the Pauli matrices. Therefore, the phase shift acquired by the wave function of a Dirac neutral particle with a permanent magnetic dipole moment is $\psi\left(x^{\mu}\right)=\mathcal{P}\,\exp\left(-i\frac{\mu}{\hbar\,c}\oint b_{\mu}\,dx^{\mu}\right)\,\psi_{0}\left(x^{\mu}\right)$. The Anandan quantum phase is an AB-type effect whose particular cases are the Aharonov-Casher effect \cite{ac} and the scalar Aharonov-Bohm effect for neutral particles \cite{zei}. Observe that we can make an analogy with the Anandan quantum phase by considering the effective vector potentials $M_{\mu}=\left(-\vec{\sigma}\cdot\vec{\mathbb{G}},\,\vec{\sigma}\times\vec{\mathbb{A}}\right)$ and $D_{\mu}=\left(-\vec{\sigma}\cdot\vec{S},\,\vec{\sigma}\times\vec{T}\right)$ from Eq. (\ref{2.11}). In the following, we show that quantum holonomies associated with the analogue of the Anandan quantum phase can be determined and, by analogy with the holonomic quantum computation \cite{zr1,vedral,vedral2}, one-qubit quantum gates can be performed.

Let us obtain the geometric quantum phase acquired by the wave function of a Dirac neutral particle. By applying the Dirac phase factor method \cite{dirac,dirac2} into the Schr\"odinger-Pauli equation (\ref{2.11}), where we can write $\psi=e^{i\phi}\,\psi_{0}$, where $\phi$ is the phase shift acquired by the wave function and $\psi_{0}$ is the solution of the Schr\"odinger-Pauli equation in the absence of fields \footnote{It is well-known, for a quantum particle with an electric charge $q$ interacting with a magnetic field $\vec{B}=\vec{\nabla}\times\vec{A}$, that the Schr\"odinger equation is given by $\hat{H}\left(\vec{r},\vec{p}-q\vec{A}\right)\psi=\mathcal{E}\psi$. As pointed out by Dirac (Refs. [46,47]), the solution $\psi$ of the Schr\"odinger equation $\hat{H}\left(\vec{r},\vec{p}-q\vec{A}\right)\psi=\mathcal{E}\psi$ can be constructed in a simple way, by writing it in terms of the wave function which is solution of the Schr\"odinger equation in the absence of fields, $\hat{H}_{0}\left(\vec{r},\vec{p}\right)\psi_{0}=\mathcal{E}\psi_{0}$, and by multiplying $\psi_{0}$ by a ``magnetic phase factor'' $\phi=\frac{q}{\hbar}\int_{\vec{r}_{0}}^{\vec{r}}\vec{A}\left(\vec{r}'\right)\cdot\,d\vec{r}'$, in the following way: $\psi=\psi_{0}\,\exp\left(\frac{iq}{\hbar}\int_{\vec{r}_{0}}^{\vec{r}}\vec{A}\left(\vec{r}'\right)\cdot\,d\vec{r}'\right)$. This is called the Dirac phase factor method \cite{dirac,dirac2}.}. In this case, then, we have that $\psi_{0}$ is the solution of the equation
\begin{eqnarray}
i\frac{\partial\psi_{0}}{\partial t}&=&\frac{1}{2m}\left[\vec{p}-i\vec{\xi}+\lambda_{1}\left(\vec{\sigma}\times\vec{T}\right)\right]^{2}\psi_{0}++\frac{\lambda_{1}}{2m}\left(\vec{\nabla}\cdot\vec{T}\right)\,\psi_{0}+\frac{\lambda_{2}}{2m}\left(\vec{\nabla}\cdot\vec{\mathbb{A}}\right)\psi_{0}\nonumber\\
[-2mm]\label{3.1}\\[-2mm]
&-&\frac{\left(\lambda_{2}\mathbb{A}\right)^{2}}{2m}\,\psi_{0}-\frac{\lambda_{2}^{2}}{2m}\,T^{2}\,\psi_{0},\nonumber
\end{eqnarray}
and general expression for the geometric phase acquired by the wave function of the neutral particle is given by
\begin{eqnarray}
\phi_{\mathrm{A}}&=&\lambda_{2}\,\oint\left[\vec{\sigma}\times\vec{\mathbb{A}}\right]\cdot d\vec{r}+\lambda_{1}\int_{0}^{\tau}\vec{\sigma}\cdot\vec{S}\,dt+\lambda_{2}\,\int_{0}^{\tau}\,\vec{\sigma}\cdot\vec{\mathbb{G}}\,dt\nonumber\\
&=&\lambda_{2}\oint\left[\vec{\sigma}\times\left(\vec{T}\times\vec{B}\right)\right]\cdot d\vec{r}-\lambda_{2}\oint\left[\vec{\sigma}\times\left(\vec{E}\times\vec{S}\right)\right]\cdot d\vec{r}\label{3.3}\\
&+&\lambda_{1}\int_{0}^{\tau}\vec{\sigma}\cdot\vec{S}\,dt+\lambda_{2}\int_{0}^{\tau}\vec{\sigma}\cdot\left(\vec{T}\times\vec{E}\right)\,dt+\lambda_{2}\int_{0}^{\tau}\vec{\sigma}\cdot\left(\vec{B}\times\vec{S}\right)\,dt.\nonumber
\end{eqnarray}

The geometric phase given in (\ref{3.3}) corresponds to the analogue of the Anandan geometric phase \cite{anan,anan2} based on a Lorentz symmetry breaking scenario defined by a tensor background $H_{\mu\nu}$. Observe that the first two integral of the second line of Eq. (\ref{3.3}) are taken when the spinor is transported in a closed path, while the three integrals of the third line of Eq. (\ref{3.3}) are taken from a time $t=0$ to $t=\tau$, where $\tau$ is the time spent by the quantum particle travelling a closed path. These integrals of the third line of Eq. (\ref{3.3}) give rise to the analogous effect of the scalar Aharonov-Bohm effect for neutral particles \cite{zei,bbs3}. Note that since the vectors $\vec{T}$ and $\vec{S}$ are considered constant vectors, therefore, the term $\lambda_{1}\left(\vec{\sigma}\times\vec{T}\right)$ does not yield any contribution to the geometric phase because it gives rise to a local term \cite{anan,anan2} as the remaining terms related to the parameters $\lambda_{1}$, $\lambda_{2}$ and $\lambda_{1}^{2}$ of Eq. (\ref{3.1}) \footnote{Observe that the term related to the parameter $\lambda_{2}^{2}$ can be neglected.}. Previous studies of the Anandan quantum phase in a Lorentz symmetry violation background \cite{belich,bbs2} has shown that the Anandan quantum phase induced by a fixed tensor/vector field background is an Abelian phase, by contrast, the Anandan quantum phase obtained in (\ref{3.3}) is a non-Abelian phase. This difference between the Abelian nature of the Anandan quantum phase in Refs. \cite{belich,bbs2} and the non-Abelian nature of the Anandan phase given in Eq. (\ref{3.3}) stems from the Lorentz symmetry violation background being defined by the coupling in (\ref{2.1}).

Now, let us consider a field configuration defined by a radial electric field produced by a uniform linear distribution of electric charges on the $z$-axis, that is, $\vec{E}=E^{1}\,\hat{\rho}=\frac{\lambda_{e}}{\rho}\,\hat{\rho}$ (where $\lambda_{e}$ corresponds to the linear density of electric charges, $\rho=\sqrt{x^{2}+y^{2}}$ is the radial coordinate, and $\hat{\rho}$ is a unit vector in the radial direction) and $\vec{S}=\left(0,\,S_{2},\,0\right)$. In this case, the Anandan quantum phase (\ref{3.3}) becomes
\begin{eqnarray}
\phi_{\mathrm{A}_{1}}&=&-\lambda_{2}\oint\left[\vec{\sigma}\times\left(\vec{E}\times\vec{S}\right)\right]\cdot d\vec{r}+\lambda_{1}\int_{0}^{\tau}\vec{\Sigma}\cdot\vec{S}\,dt\nonumber\\
[-2mm]\label{3.4}\\[-2mm]
&=&\zeta_{1}\,\sigma^{1}+\zeta_{2}\,\sigma^{2}.\nonumber
\end{eqnarray} 
where we have defined in Eq. (\ref{3.4}) the parameters: $\zeta_{1}=2\pi\,\lambda_{2}\,S_{2}\,\lambda_{e}$ and $\zeta_{2}=\lambda_{1}\,S_{2}\,\tau$. Observe that the geometric phase (\ref{3.4}) does not depend on the velocity of the Dirac neutral particle which consists in a non-dispersive geometric phase as established in Refs. \cite{disp,disp2,disp3}. We also observe that the analogue of the Anandan geometric phase given in (\ref{3.4}) is a non-Abelian phase due to the Lorentz symmetry violation background defined by a tensor field in (\ref{2.1}) in contrast to the results of Refs. \cite{belich,bbs2}, whose analogue of the Anandan geometric phase \cite{ac} is an Abelian phase due to the Lorentz symmetry violation background being defined by a fixed vector field.

Unfortunately, it is very hard to estimate a upper bound for the two Lorentz symmetry breaking terms given in the geometric phase (\ref{3.4}). Therefore, let us provide a upper bound for one of the Lorentz symmetry breaking terms, $\lambda_{1}\,S_{2}$, by supposing an experimental ability to measure geometrical phases as small as $10^{-4}\mathrm{rad}$ \cite{cimino}, then, we can affirm that the theoretical phase induced for a neutral particle cannot be larger than this value, that is, $\left|\phi_{\mathrm{A}_{1}}\right|<10^{-4}\,\mathrm{rad}$. Thereby, by considering first a null electric field and $\tau=17,8\times10^{-6}\,\mathrm{s}$ \cite{sab}. To estimate the upper bound, we establish that (for the dimensionless phase) $\phi_{\mathrm{A}_{1}}^{2}=\left[\frac{1}{\hbar}\,\lambda_{1}\,S_{2}\,\tau\right]^{2}\,<\,10^{-8}\,\mathrm{rad}^{2}$, thus, we obtain
\begin{eqnarray}
\left|\lambda_{1}\,S_{2}\right|\,<\,10^{-14}\,\mathrm{eV}.
\label{3.9}
\end{eqnarray}

Observe that the bound given in (\ref{3.9}) is obtained by a term of order $m$ which is out of SME and is evaluated in order of $10^{-23}\mathrm{GeV}$. Unfortunately, this bound is not competitive with the existing bounds in the literature $\left(10^{-28}\mathrm{GeV}\right)$ \cite{bound}.

Returning to the case given in (\ref{3.4}), we can make an analogy between the quantum holonomy associated to the Anandan quantum phase (\ref{3.4}) and the holonomic quantum computation \cite{zr1,vedral,vedral2}. The holonomic quantum computation was proposed by Zanardi and Raseti \cite{zr1} based on adiabatic cyclic evolutions with the objective of implementing quantum gates \cite{loyd} by using unitary transformation called quantum holonomies. The holonomic quantum computation is defined in the subspace spanned the eigenvectors of a family of Hamiltonian operators $\mathcal{F}=\left\{H\left(\lambda\right)=\mathbb{U}\left(\lambda\right)H_{0}\mathbb{U}^{\dag}\left(\lambda\right);\lambda\in\mathcal{M}\right\}$, where $\mathbb{U}\left(\lambda\right)$ is a unitary operator, and $\lambda$ corresponds to the control parameter that can be changed adiabatically along a loop in the control manifold $\mathcal{M}$. The action of the unitary operator $\mathbb{U}\left(\lambda\right)$ on an initial state $\left|\psi_{0}\right\rangle$ brings it to a final state $\left|\psi\right\rangle=\mathbb{U}\left(\lambda\right)\left|\psi_{0}\right\rangle$ giving rise to a quantum gate \cite{loyd}. The general expression of the action of this unitary operator is given by $\left|\psi\right\rangle=\mathbb{U}\left(\lambda\right)\left|\psi_{0}\right\rangle=e^{-i\int_{0}^{t}E\left(t'\right)\,dt'}\,\Gamma_{A}\left(\lambda\right)\left|\psi_{0}\right\rangle$, where the first terms $e^{-i\int_{0}^{t}E\left(t'\right)\,dt'}$ and $\Gamma_{A}\left(\lambda\right)$ correspond to the dynamical phase and the holonomy, respectively. The object $\mathcal{A}=A\left(\lambda\right)\,d\lambda$ is a connection $1$-form called the Mead-Berry connection 1-form \cite{tg1} and the object $A\left(\lambda\right)$ corresponds to the Mead-Berry vector potential, whose components are defined as: $A^{\alpha\beta}=\left\langle \psi^{\alpha}\left(\lambda\right)\right|\partial/\partial\lambda\left|\psi^{\beta}\left(\lambda\right)\right\rangle$. However, based on Ref. \cite{anan3}, the dynamical phase can be omitted by redefining the energy levels (for instance, by taking $E\left(0\right)=0$), then, one can study the appearance of geometric phases in any cyclic evolution of the quantum system. Recently, holonomic quantum computation has been investigated with nonadiabatic geometric phases \cite{hqc,hqc2,vedral3}, noncyclic geometric phases \cite{hqc3}, and one-qubit quantum gates associated with topological defects \cite{moraesG2,kat} have been studied for neutral particles \cite{bf9,bf30} and electrons in a crystalline solid \cite{bf10}.

In the present case, the analogy with the holonomic quantum computation \cite{zr1,vedral,vedral2} can be performed by defining first the logical states of this system as being the spin of the nonrelativistic Dirac neutral particle, that is,
\begin{eqnarray}
\left|0_{\mathrm{L}}\right\rangle=\left|\uparrow\right\rangle;\,\,\,\,\,\,\,\left|1_{\mathrm{L}}\right\rangle=\left|\downarrow\right\rangle,
\label{3.6}
\end{eqnarray}
where $\left|\uparrow\right\rangle$ and $\left|\downarrow\right\rangle$ correspond to the spin up and the spin down of the Dirac neutral particle (the spin of the neutral particle being initially polarized on the $z$-axis), respectively. This choice is justified due to the coupling of the spin of the Dirac neutral particle with the Lorentz symmetry violation background, which is manifested in the geometric phase (\ref{3.4}). Thus, the holonomy associated with the geometric phase (\ref{3.4}) in a cyclic evolution is given by
\begin{eqnarray}
\mathbb{U}\left(\zeta_{1},\zeta_{2}\right)=\exp\left(i\zeta_{1}\,\sigma^{1}+i\zeta_{2}\,\sigma^{2}\right).
\label{3.5}
\end{eqnarray}

Since the holonomy transformation (\ref{3.5}) has the sum of two non-commuting matrices into the argument of the exponential function, thus, we have that $e^{A+B}\neq e^{A}\,e^{B}$. Thus, in order to simplify the expression of the unitary operator (\ref{3.5}) acting on the logical states (\ref{3.6}), we use the Zassenhaus formula, $e^{A+B}=e^{A}\,e^{B}\,e^{-\frac{1}{2}\left[A,B\right]}\cdots$ (where $A$ and $B$ are matrices), and write the expression (\ref{3.5}) in the form \cite{bf9,bf10}: 
\begin{eqnarray}
\mathbb{U}\left(\zeta_{1},\zeta_{2}\right)\approx e^{i\zeta_{1}\,\sigma^{1}}\,e^{i\zeta_{2}\,\sigma^{2}}\,e^{i\zeta_{1}\zeta_{2}\,\sigma^{3}},
\label{3.7}
\end{eqnarray}
where we have defined the parameters $\zeta_{1}=2\pi\,\lambda_{2}\,S_{2}\,\lambda_{e}$ and $\zeta_{2}=\lambda_{1}\,S_{2}\,\tau$, and also neglected terms of order $\mathcal{O}\left(\zeta_{1}^{2}\,\zeta_{2}\right)$, $\mathcal{O}\left(\zeta_{2}^{2}\,\zeta_{1}\right)$ and higher, because we can consider these terms very small. By using the definition of of the function of a matrix, that is, $\exp{A}=\sum_{i=0}^{\infty}\frac{A^{n}}{n!}$, we can write (\ref{3.7}) in the following form:
\begin{eqnarray}
\mathbb{U}\left(\zeta_{1},\zeta_{2}\right)\approx q_{0}\,I+i\,q_{1}\,\sigma^{1}+i\,q_{2}\,\sigma^{2}+i\,q_{3}\,\sigma^{3},
\label{3.7a}
\end{eqnarray} 
where the parameters $q_{k}$ given in (\ref{3.7a}) are defined as
\begin{eqnarray}
q_{0}&=&\cos\zeta_{1}\,\cos\zeta_{2}\,\cos\zeta_{1}\zeta_{2}+\sin\zeta_{1}\,\sin\zeta_{2}\,\sin\zeta_{1}\zeta_{2};\nonumber\\
q_{1}&=&\sin\zeta_{1}\,\cos\zeta_{2}\,\cos\zeta_{1}\zeta_{2}-\cos\zeta_{1}\,\sin\zeta_{2}\,\sin\zeta_{1}\zeta_{2};\nonumber\\
[-3mm]\label{3.7b}\\[-3mm]
q_{2}&=&\cos\zeta_{1}\,\sin\zeta_{2}\,\cos\zeta_{1}\zeta_{2}+\sin\zeta_{1}\,\cos\zeta_{2}\,\sin\zeta_{1}\zeta_{2};\nonumber\\
q_{3}&=&\cos\zeta_{1}\,\cos\zeta_{2}\,\sin\zeta_{1}\zeta_{2}-\sin\zeta_{1}\,\sin\zeta_{2}\,\cos\zeta_{1}\zeta_{2}.\nonumber
\end{eqnarray}

Hence, the final step in order to define the analogy between between the quantum holonomy induced by the Lorentz-violating tensor background $H_{\mu\nu}$ and the holonomic quantum computation \cite{zr1,vedral,vedral2} is to consider the parameters $\zeta_{1}=2\pi\,\lambda_{2}\,S_{2}\,\lambda_{e}$ and $\zeta_{2}=\lambda_{1}\,S_{2}\,\tau$ (related to the Lorentz symmetry violating terms) as control parameters. The parameters $\zeta_{1}$ and $\zeta_{2}$ can be considered as control parameters in the sense that we can know or determine the values of the products $\lambda_{2}\,S_{2}\,\lambda_{e}$ and $\lambda_{1}\,S_{2}\,\tau$ previously. Therefore, knowing the values of these
parameters, we can perform an interferometry experiment with Dirac neutral particles. Thereby, applying the unitary transformation defined by Eqs. (\ref{3.7a}) and (\ref{3.7b}) to the logical states (\ref{3.6}) means that we can make a rotation on the logical states (\ref{3.6}) through the appropriate choice of the control parameters $\zeta_{1}$ and $\zeta_{2}$. By applying the holonomy transformation (\ref{3.7a}) on the logical states (\ref{3.6}) several times, then, we can perform a universal set of one-qubit quantum gates and build a toy model for the holonomic quantum computation \cite{zr1,vedral,vedral2} based on a tensor background of the Lorentz symmetry breaking. 

The phase shift yielded by the Lorentz-violating tensor background is quite small as we have seen in Eq. (\ref{3.9}). This line of investigation in the current literature searches for scenarios to establish bounds to the intensity of the background obtained by the uncertainty in measurements. This difficulty of detecting geometric phases induced by Lorentz-symmetry breaking effects in the laboratory with the present technology does not allow us to build a real model to implement the holonomic quantum computation. However, in the theoretical point of view, this work brings new discussions about geometric quantum phases in Lorentz-violation symmetry backgrounds. Quantum holonomies and non-Abelian geometric phases have not been explored in the context of the violation of the Lorentz symmetry yet. Therefore, by assuming the possibility of detecting Lorentz-violation symmetry effects, the quantum holonomy defined in Eqs. (\ref{3.7a}) and (\ref{3.7b}) allows us to produce, for instance, a superposition of states defined in Eq. (\ref{3.6}) or flip these states according to the values of the parameters  $\zeta_{1}=2\pi\,\lambda_{2}\,S_{2}\,\lambda_{e}$ and $\zeta_{2}=\lambda_{1}\,S_{2}\,\tau$. This means that we can perform one-qubit quantum gates by analogy with the holonomic quantum computation \cite{zr1,vedral,vedral2}. Hence, even though there is no experimental confirmation of Lorentz-symmetry breaking effects at present days, our proposal includes a new environment to investigate effects of the Lorentz symmetry breaking in low energies systems.

\section{conclusions}

In this work, we have shown that a possible scenario of the study of the violation of the Lorentz symmetry can be defined based on the appearance of geometric quantum phases in the wave function of a Dirac neutral particle induced by a Lorentz-violating tensor background. We have seen that analogues of the Anandan quantum phase \cite{anan,anan2} can be obtained by defining different scenarios of the Lorentz symmetry breaking induced by a tensor background $H_{\mu\nu}$. We have also seen that the geometric phase corresponding to the analogue of the Anandan quantum phase is a non-dispersive and a non-Abelian phase. The non-Abelian nature of the geometric phase stems from the Lorentz symmetry breaking scenario defined by the tensor $H_{\mu\nu}$ through the nonminimal coupling given in Eq. (\ref{2.1}). Moreover, we have seen that upper bounds for constant parameters of the Lorentz symmetry breaking can be estimated based on this geometric phase, and an analogy with the holonomic quantum computation can be made by performing one-qubit quantum gates defined by the quantum holonomies associated to the analogue of the Anandan quantum phase \cite{anan,anan2}.

We would like to thank CNPq (Conselho Nacional de Desenvolvimento Cient\'ifico e Tecnol\'ogico - Brazil) for financial support.

\end{document}